\documentclass{aa}
\usepackage{graphicx}

\usepackage{natbib}
\begin{document}

\title{Spectroscopic confirmation of a cluster of galaxies at $z=1$ in 
the field of the gravitational lens MG2016+112
\thanks{Based on observations made with the Nordic Optical Telescope,
operated on the island of La Palma jointly by Denmark, Finland, Iceland, 
Norway, and Sweden, in the Spanish Observatorio del Roque de los Muchachos 
of the Instituto de Astrofisica de Canarias. Based on observations 
with the Canada-France-Hawaii Telescope at Mauna Kea, Hawaii, USA.}
}
\authorrunning{G. Soucail et al.}
\titlerunning{Spectroscopic observations of MG2016+112}

\author{
G. Soucail\inst{1} 
\and J.-P. Kneib\inst{1} 
\and A. O. Jaunsen \inst{2,3}
\and J. Hjorth \inst{4,5,3}
\and M. Hattori \inst{6} 
\and T. Yamada \inst{6}
   }

   \offprints{Genevi\`eve Soucail, \email{soucail@ast.obs-mip.fr}}

   \institute{Observatoire Midi-Pyr\'en\'ees, Laboratoire d'Astrophysique, 
    UMR 5572, 14 Avenue E. Belin, F-31400 Toulouse, France 
    \and Institute of Theoretical Astrophysics, University of Oslo, 
         Pb.\ 1029, Blindern, N--0315 Oslo, Norway
    \and Centre for Advanced Study, Drammensvn.~78, N--0271 Oslo, Norway
    \and Astronomical Observatory, University of Copenhagen, 
         Juliane Maries Vej 30, DK--2100 Copenhagen \O, Denmark
    \and NORDITA, Blegdamsvej 17, DK--2100 Copenhagen \O, Denmark
    \and Astronomical Institute, T\^ohoku University, Aoba Aramaki,
     Sendai 980--77, Japan 
    }
   
\date{Received June 27, 2000 / Accepted December 13, 2000}

\abstract
{We present new optical data on the cluster AX~J2019+1127 identified by
the X-ray satellite ASCA at $z\sim 1$ \cite{hattori97}. The data
suggest the presence of a high-redshift cluster of galaxies
responsible for the large separation triple quasar MG2016+112. Our
deep photometry reveals an excess of $z\sim 1$ galaxy candidates, as
already suspected by Ben\'\i tez et al.\ \cite*{benitez99}. Our
spectroscopic survey of 44 objects in the field shows an excess of 6
red galaxies securely identified at $z \sim 1$, with a mean redshift
of $z =1.005\pm 0.002$. We estimate a velocity dispersion of $\sigma =
771^{+430}_{-160}$ km\,s$^{-1}$ based on these 6 galaxies and a V-band
mass-to-light ratio of $215^{+308}_{-77} \ h_{50} \ (M/L_V)_\odot$.  Our
observations thus confirm the existence of a massive structure acting
as the lens, which explains the unusual configuration of the triple
quasar. Hence, there is no need to invoke the existence of a
``dark cluster'' to understand this lens system.
\keywords{cosmology: observations -- dark matter -- 
galaxies: clusters: general -- 
galaxies: clusters: individual: MG2016+112 -- 
gravitational lensing -- X-rays: galaxies}
}

\maketitle

\section{Introduction}
Gravitational lensing is an efficient way to probe the mass content of
the Universe. It acts on many scales, from dark stellar-mass objects
to the largest massive structures \cite{mellier99}. Some of the most
spectacular examples of this phenomenon are the multiple quasar
systems already known for 20 years \cite{walsh79}. However, in most
cases simple lens models cannot reproduce the image geometry and flux
ratios, unless an additional external component such as {\it external
shear} is included in the modeling \cite{keeton97}. Furthermore in the
cases of large-separation multiple quasars, the mass-to-light ratio of
the main galaxy lens is generally much higher than what is usually
expected for a galaxy. As an example, the double quasar Q0957+561,
with a separation of 6\arcsec, is well explained with a deflector
corresponding to a massive and bright galaxy combined with the
additional effect of a cluster of galaxies centered on it (Bernstein
\& Fisher 1999\nocite{bernstein99}, and references therein).  This
cluster was indeed spectroscopically confirmed at the redshift of the
main galaxy deflector \cite{angonin94}. In general, it is becoming
evident that the likely explanation of the external shear
necessary to understand the wide separation lensing configurations is
the existence of a group/cluster near the line of sight of these
multiple quasars. In particular, cluster mass distributions have been
detected in several other wide-separation multiple quasars such as the
so-called ``Cloverleaf'' \cite{kneib98} and RX~J0911+05
\cite{burud98,kneib00}.

A mysterious system in this respect is the triple quasar MG2016+112
($z=3.26$). It was discovered by Lawrence et al.\ \cite*{lawrence84}
and studied in more detail by Schneider et al.\
\cite*{schneider85,schneider86a} in the optical and by Garret et al.\
\cite*{garret94,garret96} in the radio. The third image named C
presents a complex structure, suggesting some differential
magnification of the source within each image. Deep imaging revealed
the existence of a giant elliptical galaxy at $z \simeq 1.01$ named D
and located between the 3 images. It is identified as partly
responsible for the multiple lensing of the quasar, although its mass
is insufficient to explain the complex nature of the lensed images of
MG2016+112. Detailed modeling requires another source of lensing mass,
such as the mass of a rich cluster, centered on galaxy D
\cite{narasimha87}. In order to reveal the exact nature of the lens,
deep X-ray searches have been attempted to probe hot intra-cluster
gas. A positive X-ray detection of a cluster-like emission was
obtained, centered on MG2016+112 \cite{hattori97}. The X-ray
characteristics (notably an emission line at $3.49^{+0.15}_{-0.13}$
keV and the extended nature of the emission) indicate that
AX~J2019+1127 is an X-ray cluster of galaxies at $z \sim 1$.  The
X-ray temperature related to this emission gives $k$T = $8.6
^{+4.2}_{-3.0}$ keV and an X-ray luminosity L$_X = 8.4^{+2.4}_{-1.7} \
h_{50}^{-2}$ erg s$^{-1}$ in the 2--10 keV band, quite compatible with
the L$_X$--T$_X$ relation for clusters of galaxies
\cite{markevitch98}.  The mass derived from the X-ray analysis is
about $3\times 10^{14} \ h_{50}^{-1}$ M$_\odot$ within 500
$h_{50}^{-1}$ kpc, a typical value for rich clusters of galaxies with
velocity dispersion of $\sim 1200$ km\,s$^{-1}$ (assuming the observed
$\sigma-T_X$ relation, Girardi et al. 1996\nocite{girardi96}). A
re-analysis of the HRI/ROSAT imaging \cite{benitez99} seems to
slightly decrease the expected cluster mass, without rejecting its
detection. In addition, deep near-IR imaging of the field was used to
identify a `red sequence' in the color-magnitude diagramme, presumably
due to early-type galaxies in the cluster.

Recently, Clowe et al.\ \cite*{clowe00} revisited the Ben\'\i tez et
al.\ optical data, complemented with ultra-deep Keck R-band
images. Again a sequence of cluster galaxies was clearly identified
and its spatial distribution analyzed. More interesting is their weak
lensing analysis of the faint galaxies from the R-band image, which
seems to give a signal in the field. Its center appears offset about
1\arcmin\ North of MG2016+112 itself and its mass is consistent with
the mass inferred from X-ray data. The authors claim they can rule out
a mass centered exactly on MG2016+112, at a $2\sigma$
level. Complementary data are clearly required to better quantify
these preliminary results of relatively low significance.

In this paper, we present new spectroscopic observations in the field
of this peculiar lens, aimed at identifying the lensing cluster. 
Observations and data reduction are described in Section 2,
the results are presented in Section 3.  Section 4 presents a
preliminary analysis of the lensing cluster identified at $z \simeq
1$. Section 5 gives some conclusions and prospects in the
understanding of this object.

Throughout the paper, we use a Hubble constant of H$_0 = 50 \ h_{50}$ km
s$^{-1}$ Mpc$^{-1}$, with $\Lambda = 0$ and $\Omega_0 = 1$. At the cluster
redshift ($z=1$), 1\arcsec\ corresponds to 8.52 $h_{50}^{-1}$ kpc
and 1 $h_{50}^{-1}$ Mpc covers 2\arcmin.

\section{Observations and data reduction}

\subsection{Photometry}
The imaging data were obtained in R and I-bands in June 1996 at the
Nordic Optical Telescope, using the HiRAC instrument with a LORAL
2k$\times$2k CCD, giving a pixel size of 0\farcs107.  The data
reduction was performed in the IRAF\footnote{IRAF is distributed by
NOAO, which are operated by AURA, Inc., under cooperative agreement
with NSF.} environment.  Observations of M92 at various airmasses
provided a calibration to the Landolt system \cite{landolt92} based on
CCD photometry of the field (L. Davis, private communication). The
zero-point, color coefficient and extinction was then computed using
the {\tt PHOTCAL} package. In the relatively crowded field of
MG2016+112 bright foreground stars with a significant PSF tail
extending to many tens of arc-seconds hampered conventional photometry
of the much fainter galaxies of interest. Moreover, PSF subtraction of
point sources in the combined image proved difficult due to
interpolation artifacts during the image mapping to a common
coordinate system.  Instead the bright unsaturated stars were removed
by PSF subtraction on the individual images.  To compensate for
possible PSF variations in the field we experimented with both a
constant and second order varying PSF. No significant differences were
detected, however, and the constant PSF was therefore used. The
spatial image transformations determined from the pre-PSF subtracted
images were then applied and combined to get the final R and I
images. The FWHM of unsaturated stars in these images were found to be
0\farcs75 and 0\farcs69 in the R and I-bands, respectively.

Galaxy photometry was computed using the MAG\_BEST estimate from
SExtractor \cite{bertin96} and transformed to the standard Landolt
system using the derived photometric transformation. The photometric
catalogue was not corrected for Galactic extinction although for this
particular field, the correction could be as high as 0.5 magnitude,
depending on the filter.
Finally conversion to AB magnitudes was done using coefficients
(c$_{\rm R}=0.169$, c$_{\rm I}=0.432$) given in Fukugita
et~al.~\cite*{fukugita95}. A total of 7 R-band images ($4500$ sec) and
10 I-band images ($6000$ sec) were combined reaching 3$\sigma$ sky
limits of $25.4$ and $25.3$ within the seeing disk (1.5$\times$FWHM),
respectively. A significant fraction of faint (cluster member)
galaxies are expected to remain undetected owing to the strong halos
of saturated stars in the field and the fact that the photometry of
the field is not extremely deep.

\subsection{Spectroscopy}
Spectroscopic data were acquired on the nights August 2--4, 1997 at
the Canada-France-Hawaii Telescope with the OSIS multi-object
spectrograph \cite{lefevre94}. A tip-tilt correction is introduced
thanks to the guiding on a selected star in the field. The seeing is
then significantly improved ($\sim 0\farcs5$ measured on our images),
increasing the detectivity of the instrument for faint object
spectroscopy. We used the $2048 \times 2048$ STIS2 thinned CCD, with
pixels of 21 $\mu$m (or 0\farcs151 on the sky) covering a field of
view of $3\farcm6 \times 3\farcm6$. We also used the R150 grism, giving
a useful spectroscopic signal from 4500 to 9000 \AA\ and a dispersion
of 3.7\AA /pixel. The slits were 10\arcsec\ long and 1\farcs0 in
width, giving a final instrumental resolution of 18\AA .  Two masks
were punched with 19 slits and 18 slits respectively, and observations
were obtained for 21600 seconds for Mask1 (6 exposures) and 23400
seconds for Mask2 (7 exposures) in good weather conditions with seeing
ranging from 0\farcs5 to 0\farcs8.  The spectroscopic targets were
selected from the photometric catalogue, with the following criteria:
$22<I_{AB}<23.3$ and $(R-I)_{AB} > 0.5$, with a careful selection of
resolved objects. These criteria were optimized to avoid a strong
contamination of the sample by the faint stars crowding at this low
galactic field, and to increase the chance of selecting high redshift
galaxies.

\begin{figure*}
\centerline{\includegraphics[width=0.9\textwidth,angle=-90]{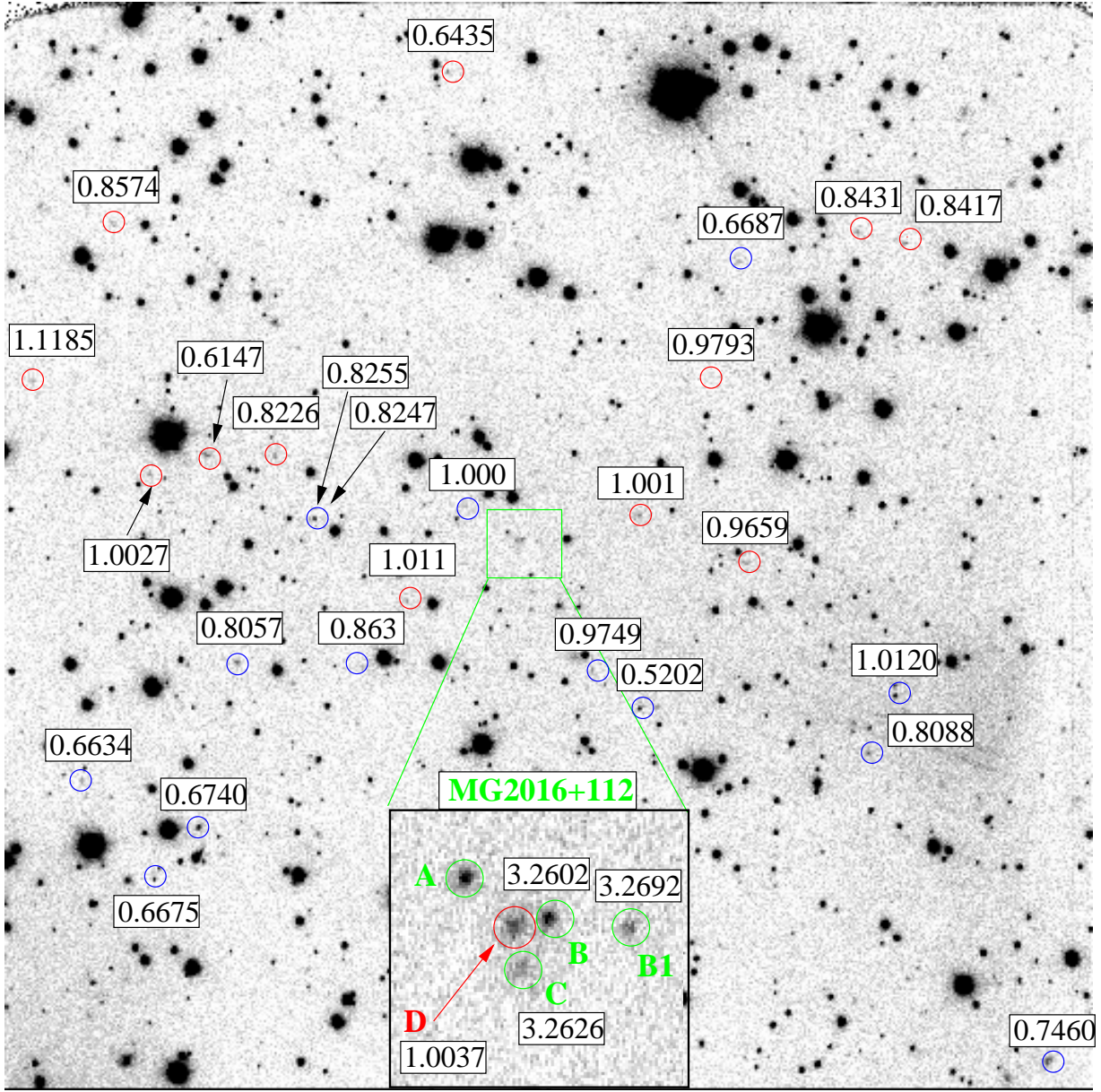}}
\caption{I image of the field of MG2016+112 from the NOT, with the
location of the spectroscopically identified objects. The field of
view of the image corresponds to 3\farcm5 $\times$ 3\farcm5, or a
physical size of $1.75 \ h_{50}^{-1}$ Mpc $\times 1.75 \ h_{50}^{-1}$
Mpc at the cluster redshift. The seeing of this image is
0\farcs7. North is up, East is left.}
\label{ident-field}
\end{figure*}

Data were reduced with the {\tt MULTIRED} package
\cite{lefevre95}. Spectra were bias-subtracted, flat-fielded and
extracted. The wavelength calibration gave typical internal errors of
$\sim 0.3$ \AA .  The spectra were finally rebinned to a dispersion of
6\AA /pixel, better matched to the resolution of the instrument. We
encountered some difficulties with the flux calibration in the red end
of the spectra because no order-separating filter was used. So
contamination by the second order contribution was significant above
7500--8000\AA , giving lower confidence to the flux calibration above
this limit.

\begin{table*}
\caption[]{Catalogue of the observed objects in the field of
MG2016+112. The absolute coordinates of the objects are computed in
the FK5 system. The magnitudes are computed in the AB system and are
not corrected for galactic extinction. Object labelled M2\_1 has no
magnitude because it falls out of the NOT field of view. Objects
M2\_13a and M2\_13b are not deblended in the photometric catalogue,
the magnitude given in the table corresponds to the sum of the two
objects.}
\label{tab-z}
\begin{flushleft}
\begin{tabular}{lrrcccl}
\hline\noalign{\smallskip}
 Name\_id & $\alpha_{2000} $ & $\delta_{2000} $ & \quad
 $I_{AB}$ & $(R-I)_{AB}$ & z & Comments \\
\noalign{\smallskip}
\hline\noalign{\smallskip}
      M1\_5 & 20:19:12.90 & +11:28:11.5 & 22.20 & 1.11 & 0.8417 & 
[OII](em) - bluer continuum than M1\_6 \\
      M1\_6 & 20:19:13.51 & +11:28:13.7 & 22.69 & 1.68 & 0.8431 & 
[OII](em), [OIII](em) \\
      M1\_7 & 20:19:15.02 & +11:27:08.8 & 22.16 & 1.17 & 0.9659 & 
[OII](em), H$\delta$ \\
      M1\_8 & 20:19:15.80 & +11:27:44.4 & 22.80 & 1.23 & 0.9793 & 
[OII](em), H$\delta$, H$\gamma$ ? \\
      M1\_9 & 20:19:16.43 & +11:27:18.2 & 22.14 & 1.73 & 1.001  & 
4000\AA\ break, G \\
    M1\_11a & 20:19:17.79 & +11:27:13.1 & 23.29 & 1.36 & 3.2692 & 
{\bf Object B1} -- Ly$\alpha$(em) \\
    M1\_11b & 20:19:17.99 & +11:27:13.4 & 22.78 & 0.45 & 3.2602 & 
{\bf Radiosource B }\\
    M1\_11c & 20:19:18.07 & +11:27:13.2 & 22.41 & 1.86 & 1.0037 & 
{\bf Object D} -- [OII](em), H\&K, 4000\AA\ break, G \\
     M1\_12 & 20:19:18.98 & +11:28:45.0 & 23.03 & 0.64 & 0.6435 & 
[OII](em), H\&K, H$\delta$, [OIII](em) \\
     M1\_13 & 20:19:19.54 & +11:27:01.4 & 22.65 & 1.57 & 1.011  & 
[OII](em), H$\varepsilon$, 4000\AA\ break, H$\delta$? \\
     M1\_15 & 20:19:21.30 & +11:27:29.9 & 22.64 & 1.61 & 0.8231 & 
H\&K, 4000\AA\ break, G \\
     M1\_16 & 20:19:22.24 & +11:27:30.2 & 21.34 & 0.91 & 0.6147 & 
[OII](em), H$\delta$, [OIII](em) \\
     M1\_17 & 20:19:22.98 & +11:27:26.4 & 22.16 & 1.77 & 1.0027 & 
weak [OII](em), H\&K, 4000\AA\ break \\
     M1\_18 & 20:19:23.47 & +11:28:15.6 & 22.34 & 0.98 & 0.8574 & 
 [OII](em), H$\delta$, H$\gamma$ \\
     M1\_19 & 20:19:24.55 & +11:27:44.5 & 22.76 & 0.88 & 1.1185 & 
[OII](em), H\&K, 4000\AA\ break \\
      M2\_1 &   ---   &    ---  &  ---  & ---  & 0.3143 &
H\&K, 4000\AA\ break, G \\
      M2\_3 & 20:19:10.98 & +11:25:31.0 & 21.01 & 1.01 & 0.7460 & 
[OII](em), H\&K, 4000\AA\ break, H$\delta$, H$\gamma$ \\
      M2\_4 & 20:19:12.87 & +11:26:39.4 & 22.92 & 1.15 & 1.0120 & 
[OII](em), K, H$\delta$, H$\gamma$ \\
      M2\_5 & 20:19:13.37 & +11:26:31.5 & 22.11 & 1.53 & 0.8088 & 
[OII](em), H\&K, 4000\AA\ break, H$\delta$ \\
      M2\_7 & 20:19:15.10 & +11:28:07.8 & 22.43 & 0.58 & 0.6687 & 
[OII](em), [OIII](em) \\
      M2\_8 & 20:19:16.19 & +11:26:40.5 & 22.18 & 0.32 & 0.5202 & 
[OII](em), H\&K, G, [OIII](em) \\
      M2\_9 & 20:19:17.06 & +11:26:47.8 & 23.24 & 1.42 & 0.9749 & 
4000\AA\ break, G ? \\
     M2\_10 & 20:19:18.05 & +11:27:11.6 & 22.60 & 0.94 & 3.2626 & 
{\bf Radiosource C} \\
     M2\_11 & 20:19:18.75 & +11:27:19.8 & 22.36 & 1.59 & 1.000  & 
MgII(2800), H\&K, 4000\AA\ break, H$\delta$, H$\gamma$ \\
     M2\_12 & 20:19:20.14 & +11:26:48.4 & 22.19 & 0.90 & 0.863  &
4000\AA\ break, H$\delta$ \\
    M2\_13a & 20:19:20.78 & +11:27:17.7 & 22.18 & 1.40 & 0.8247 & 
[OII](em) ?, H\&K, 4000\AA\ break, G \\
    M2\_13b & 20:19:20.83 & +11:27:17.4 & ---   & ---  & 0.8255 & 
[OII](em), [OIII](em) \\
     M2\_14 & 20:19:21.82 & +11:26:49.5 & 22.50 & 0.75 & 0.8057 & 
[OII](em), 4000\AA\ break, H$\delta$ \\
     M2\_15 & 20:19:22.32 & +11:26:17.1 & 21.74 & 1.28 & 0.6740 & 
[OII](em), H\&K, 4000\AA\ break, H$\delta$, G, [OIII](em) \\
     M2\_16 & 20:19:23.28 & +11:26:06.7 & 20.74 & 1.17 & 0.6675 & 
[OII](em), H\&K, 4000\AA\ break, G \\
     M2\_17 & 20:19:23.88 & +11:26:26.1 & 22.80 & 1.33 & 0.6634 & 
H\&K, 4000\AA\ break, G \\
\noalign{\smallskip}
\hline
\end{tabular}
\end{flushleft}
\end{table*} 

Finally, the redshift measurement of the spectra was done, using the
{\tt RVSAO2.0} package \cite{kurtz98} and correlating the observed
spectra with a set of stellar and galactic templates. Because of the
low S/N for most of the spectra, a visual check of the spectral
identification was done carefully as well, in order to help the
redshift measurement of the less significant spectra.

\begin{figure}
\centerline{\includegraphics[width=0.49\textwidth]{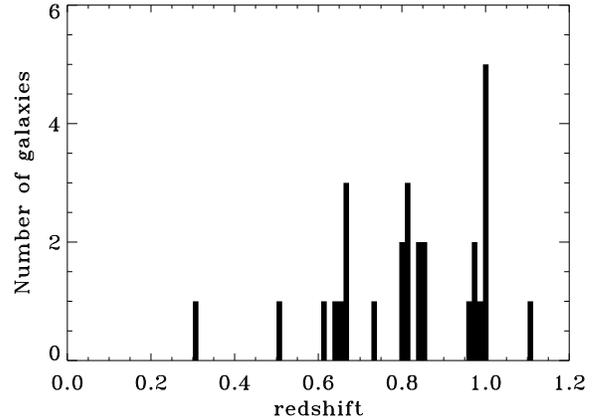}}
\caption{Redshift Histogramme of our spectroscopic sample. The bin
width was adjusted to correspond to 2000 km s$^{-1}$ at the cluster
redshift.}
\label{histo-z}
\end{figure}

Among the 44 spectra extracted (in some of the 37 slits, more than 1
spectrum was extracted), 10 correspond to stars and 3 remain
unidentified due to a poor S/N ratio, but these are most probably
non-stellar. For the 31 remaining spectra the redshift identification
is presented in Table \ref{tab-z} and in Fig.~\ref{ident-field}. The
redshift distribution of these 31 objects is shown in
Fig.~\ref{histo-z}.  The sample is far from complete in magnitude so
no conclusion can be drawn on the redshift distribution of the
galaxies along this line of sight. An excess of galaxies around $z=1$
is, however, clearly seen and corresponds to the redshift of galaxy D
already identified in Schneider et al.\ \cite*{schneider86a}. Two
other concentrations are suspected which are also spatially coherent
at $z \simeq 0.66$ and $z \simeq 0.82$, with a somewhat lower
significance.

\section{The lensed source MG2016+112}
\begin{figure}
\centerline{\includegraphics[width=0.5\textwidth]{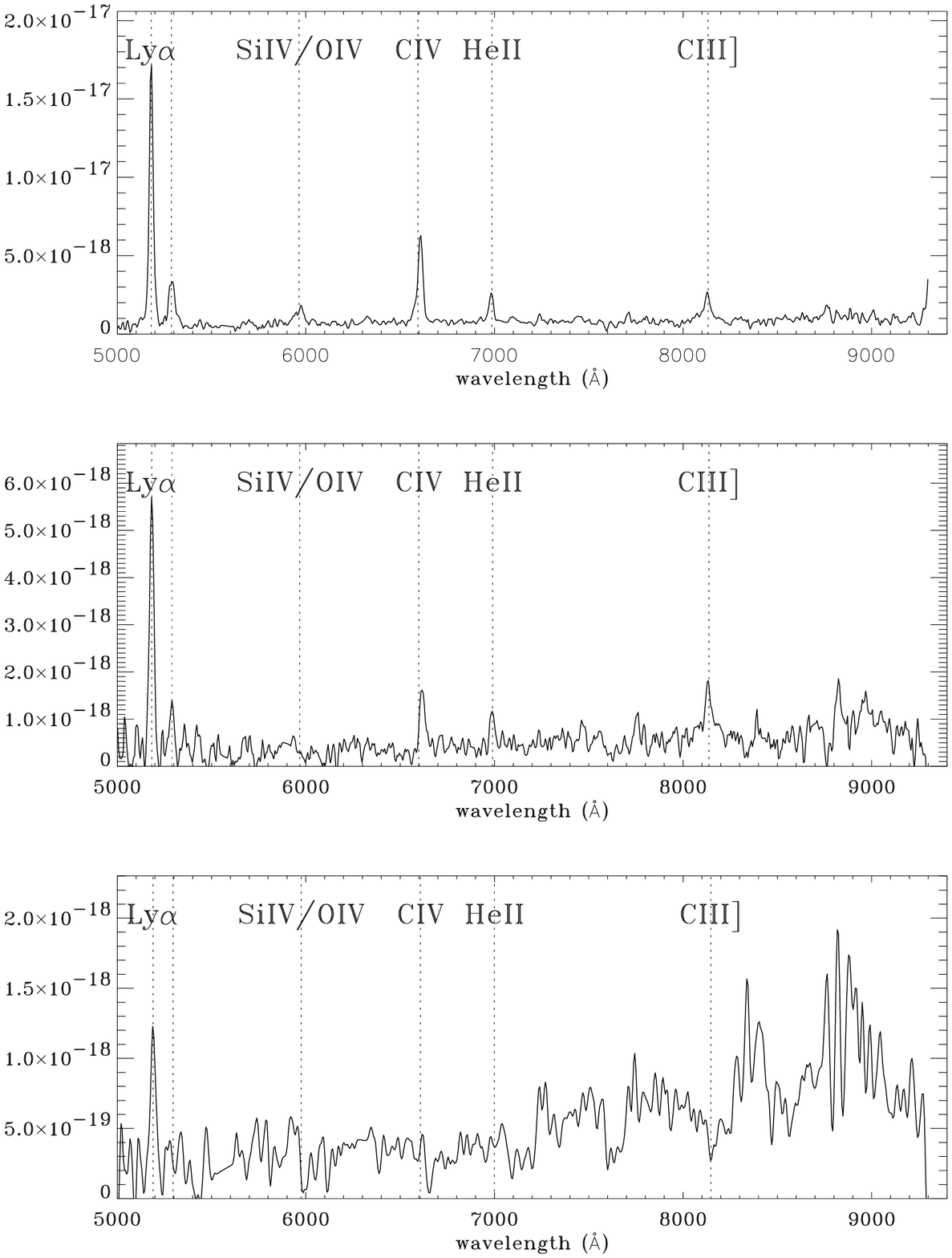}}
\caption{Spectra of the objects labeled B (top) C (middle) and B1
(bottom). The signal above 7500 \AA\ is hardly significant for B1 and
may be due to sky residuals. Most of the emission lines of B and C
analyzed in Yamada et al.\ (2000) are indicated.}
\label{b1}
\end{figure}

Most of our results on the multiple images of the radio source
MG2016+112 are presented in Yamada et al.\ \cite*{yamada00} who study
in detail the emission-line properties of the images B and C. Note
that this is the first spectrum of C observed in the optical
(Fig. \ref{b1}), after its identification as a third image by its
Ly$\alpha$ emission \cite{schneider86a} and its radio emission
\cite{garret94}. The higher resolution of our spectrograph compared to
the spectrum of B presented in Schneider et al.\ \cite*{schneider85}
allows to resolve the main emission lines and then to study the
physics of the emitting regions in the source.

Schneider et al. \cite*{schneider86a} claimed that 2 objects, labeled
A1 and B1, detected close to the triple radio-source, were at the same
redshift as the lensed source. They identified them from a narrow-band
image of the field, centered on Ly$\alpha$ redshifted to 3.26. A close
look at our deep I and R images confirms the detection of B1 but not
A1. B1 was positioned in one slit of Mask 2 and its spectrum
(Fig. \ref{b1}) confirms its redshift at $z= 3.269$, similar but
different from MG2016+112. Note however that the emission-line
properties of B1 are quite different from those of B, with only
Ly$\alpha$ in emission. SiIV and CIV may be detected in absorption but
with a lower confidence. If true, these lines are representative of a
young stellar component in the source, rather than an active nucleus.

\section{The lensing cluster at $z \simeq 1$} 
\begin{figure*}
\centerline{\includegraphics[angle=90,width=\textwidth]{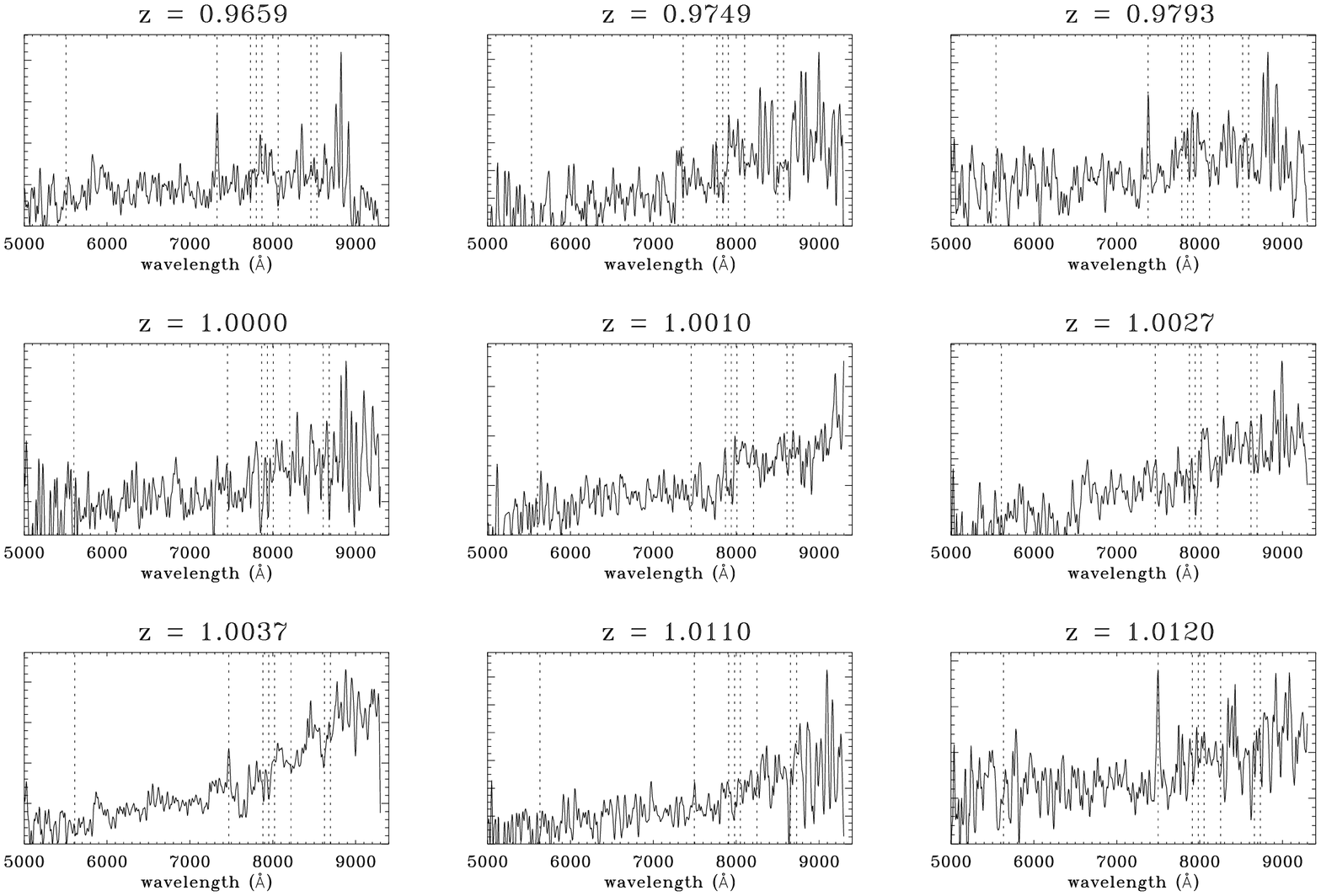}}
\caption{Spectra of the 9 galaxies identified as cluster members. The
wavelength scale is in \AA . The spectra are flux calibrated (in erg
s$^{-1}$ cm$^{-2}$ \AA $^{-1}$) and smoothed with a window
corresponding to the resolution of the spectrograph.  The main
characteristic spectral lines (MgII $\lambda$2800, [OII]
$\lambda$3727, the H\&K CaII lines, the 4000 \AA\ break, H$\delta$,
the G-band and $H\gamma$) redshifted for each object are indicated as
dashed lines.  Galaxy D which is at the center of the lens MG2016+112
is at redshift 1.0037.}
\label{cluster-z1}
\end{figure*}

In our spectroscopic sample, 9 objects have a redshift $z\sim 1$, with
8 new spectroscopic identifications (Fig. \ref{cluster-z1}). This
confirms that a coherent gravitational structure is associated with
object D, previously suspected as the main lens \cite{hattori97}. Not
surprisingly, these objects are among the reddest ones in Table
\ref{tab-z}. This justifies {\em a posteriori} the color selection we
introduced in the mask production. Moreover, if one concentrates on
the objects selected in a 0.5 $h_{50}^{-1}$ Mpc radius around galaxy
D, we note that all galaxies are at $z\sim 1$ except one at $z=0.52$.
As there are not many other red galaxies in this region, we may have
identified most of the brighter cluster galaxies. In this respect deep
IR imaging should help to identify fainter cluster members, because
the foreground contamination is less important in the near-IR.

From multicolor photometry only, Ben\'\i tez et al.\ \cite*{benitez99}
followed by Clowe et al.\ \cite*{clowe00} attempted to isolate cluster
galaxies from their colors in the red and near-infrared. Of course
this selection procedure favors old-type galaxies and avoids strong
star-forming ones which are more likely mixed with field galaxies. We
compared their sample with our spectroscopically confirmed cluster
members and found a good agreement with their selected objects. In
particular we confirm the cluster membership of objects B5 ($\equiv$
M1\_9, $z=1.001$), B6 ($\equiv$ M1\_13, $z=1.011$) and B2 ($\equiv$
M2\_11, $z=1.000$), while we find a lower redshift $z = 0.8231$ for
object labeled \# 3 in Clowe et al.\ \cite*{clowe00}. Note that their
identification was done after our preliminary results presented in the
CFHT Inofrmation Bulletin (report \# 38, 1998), so their sequence may
not be quite independent from our first results.

Among the 9 galaxies at $z\sim 1$, 4 galaxies (+ 2 more uncertain
ones) show the [OII] 3727\AA\ line in emission, indicating active star
formation in these galaxies. This may be a sign of the relative youth
of this cluster or sign of interaction between the galaxies and the
intra-cluster medium. Note that similar trends were observed in a few
other high redshifts clusters \cite{postman98}. To address in more
detail the question of the evolution of galaxies and their spectral
content in high redshift galaxies, a large sample of cluster members
should be analyzed spectroscopically as well as through their optical
and near-IR colors, including morphological information. This kind of
data is now within the reach of 8--10m class telescopes equiped with
optical and near-IR instruments.

From a dynamical point of view, we can derive a rough estimate of the
velocity dispersion, although with small numbers it is difficult to
measure accurate properties. If we keep the total sample of 9
galaxies, we find a mean redshift of $\bar{z} =0.995$ and a velocity
dispersion of $\sigma_{los} = 2510^{+960}_{-450}$ km s$^{-1}$.  The
maximum difference is $\Delta v = 1/2 \ (v_{max} - v_{min}) = \pm
3500$ km s$^{-1}$. These values are rather unrealistic for a well
defined dynamical structure. But among our 9 galaxies, 3 are isolated
around a redshift $z=0.97$, while the 6 others are spread around
$z=1$. We suspect the first 3 galaxies may belong to a small structure
such as a group, spatially and dynamically close to the main one.  But
we are well aware that these assumptions are quite unsecure because of
the small numbers.  Considering then the 6 galaxies in the redshift
range $[1.0;1.012]$ we find a mean redshift of $\bar{z} =1.005$, a
velocity dispersion of $\sigma_{los} = 771^{+430}_{-160}$ km s$^{-1}$
and a velocity range: $\Delta v = \pm 900$ km s$^{-1}$. These values
make more sense in view of the X-ray observations, although the error
bars are still quite large.

Using the virial theorem, we can estimate the mass of this system
within the harmonic radius $R_h$ defined as \cite{nolthenius87}:
\begin{equation}
\displaystyle
R_h = D_A(\bar{z}) {\pi\over 2}{N_m(N_m-1)\over 2}
\left(\Sigma_i\Sigma_{j>i}\theta_{ij}^{-1}\right)^{-1},
\end{equation}
where $\theta_{ij}$ is the angular distance between galaxies $i$ and
$j$, $N_m$ is the number of cluster members, and $ D_A(\bar{z})$ is
the angular diameter distance at the mean cluster redshift $\bar{z}$.
The cluster virial mass can then be estimated as
\begin{equation}
M = { 6 \sigma^2 R_h \over G}.
\end{equation}
With our sample of 6 galaxies, we find an harmonic radius of $R_h =
348 \ h_{50}^{-1}$ kpc and a virial mass of $M = 2.8 ^{+4.0}_{-1.0} \
10^{14} \ h_{50}^{-1}$ M$_\odot$. The error bars quoted in the virial
mass correspond to the errors in the velocity dispersion measure
only. They do not include uncertainties due to the harmonic radius
although they may be significant, because of our very small sample of
cluster members. As quoted by Carlberg et al.\ \cite*{carlberg96}, the
use of the harmonic radius may underestimate the virial mass
determination because $R_h$ is highly sensitive to close pairs and
rather noisy.

We can also compare our dynamical mass with the galaxy
distribution to get an estimate of the mass-to-light ratio of the
structure. Our own measure of the luminosity of the 6 cluster
members, corrected the same was as Ben\'\i tez et al.\
\cite*{benitez99} [k-correction of 3.42 in V and 1.12 in I,
average color index $V-I=3.2$ for the galaxies, average
extinction correction $A_I=0.44$ magnitude] gives a total
luminosity of $L_V^{gal} = 7.6 \times 10^{11} \ h_{50}^{-2}
L_{V\odot}$ for these galaxies. It is about 50\%\ higher
than the luminosity of their 9 cluster member candidates and
60\%\ smaller than their extrapolated total magnitude. A more
accurate luminosity function determination is necessary for a
correct estimate of the total luminosity, and the M/L
ratio. Following the reasonable total luminosity function of the
cluster galaxies estimated by Ben\'\i tez et al.\
\cite*{benitez99} ($L_V^{all} = 1.3
\times 10^{12} \ h_{50}^{-2} L_{V\odot}$) we find a mass-to-light ratio
of
\[ (M/L_V) = 215^{+308}_{-77} \ h_{50} \ (M/L_V)_\odot \]
This value can be compared with two other estimates proposed for this
cluster. From the X-ray emission and the measurements of Hattori et
al.\ \cite*{hattori97} we find a value $M_X/L_V = 224 \ h_{50} \
(M/L_V)_\odot$ within a radius of 400 $h_{50}^{-1}$ kpc. Note that an
error of a factor two occurs in the estimate of $M_X/L_V$ in Ben\'\i
tez et al.\ \cite*{benitez99}, due to an error in the conversion
between $h_{50}$ and $h_{100}$ units. From an extrapolation of their
lens model they also find a lensing $M/L_V$, within a radius of 400
$h_{50}^{-1}$ kpc of 186 $h_{50} \ (M/L_V)_\odot$. Our determination
falls within the same range as these two estimates. All of them are
based on the cluster luminosity estimated by Ben\'\i tez et al.\ which
is most probably underestimated because of the difficulties in
selecting cluster members in the optical in this low galactic latitude
field. In addition, no evolution corrections are included, which at
such high redshift would increase the $M/L$ ratio, typically by a
factor of 2 \cite{smail97}. Better estimates will come from near-IR
imaging where both effects are strongly reduced. In all cases, the
$M/L$ value proposed for MG2016+112 is slightly higher than other
accepted values found in other high redshift clusters: Smail et al.\
\cite*{smail97} find an average $M/L_V = 90 \ h_{50} \ (M/L_V)_\odot$
within 800 $h_{50}^{-1}$ kpc for their sample of clusters in the
redshift range [0.2--0.5], while Carlberg et al.\ \cite*{carlberg96}
have a median value of $M/L_r = 143 \ h_{50} \ (M/L_V)_\odot$ for the
CNOC sample. For a more distant cluster ($z \simeq 1$), Deltorn et
al.\ \cite*{deltorn97} find a slightly higher value of $M/L_B = 200 \
h_{50} \ (M/L_B)_\odot$ within a radius of 400 $h_{50}^{-1}$ kpc
from 11 cluster members. In
any case, MG2016+112 is not a ``dark cluster'' anymore with a $M/L$
ratio as high as 1000 as suspected initially when most of the light
was concentrated on galaxy D.

\section{Conclusions}
We have presented a photometric and spectroscopic survey of the field
around the triple lens MG2016+112. Our observations confirm the
identification of a massive cluster at $\bar{z}=1.005$
\cite{hattori97}.  With the identification of 6 secure cluster members
we have estimated the velocity dispersion $\sigma_{los}\sim
800$km\,s$^{-1}$ and used this to infer the virial mass associated
with the cluster. Although we are well aware of the limitations due to
the small number statistics of our sample in the analysis of the
cluster mass, we also tried to compare with other estimates such as
the mass inferred from the X-ray analysis or a weak lensing one. All
the M/L ratios derived that way are consistent with each other. The
value found ($M/L = 215^{+308}_{-77} \ h_{50} \ (M/L)_\odot$ for the
dynamical analysis) is also consistent with what is found in other
high redshift clusters, although we cannot yet deduce any evolution in
this ratio.  But clearly the cluster around MG2016+112 is not a dark
cluster anymore.

To overcome most of the observational difficulties due to the low
galactic latitude of the field, we underline the need of deep
near-infrared imaging to increase the density contrast of $z=1$
galaxies clustered around the lensed source and to reduce the
uncertainties in the luminosity function of these galaxies. A deeper
spectroscopic follow-up with a spectrograph on a 8--10m telescope will
also increase the identification of cluster members and a better
dynamical analysis. Furthermore, deep X-ray observations with the 2
new X-ray satellites, {\it Chandra} and {\it XMM-Newton}, will also
increase the accuracy in the cluster X-ray emission (exact location of
the center, good determination of the mass profile, better measure of
the temperature and Fe abundance, etc.), leading to a more accurate
cluster mass measure.

As the number of clusters of galaxies at redshift $z > 1$ is still
small it is valuable to focus on their details to better understand
the physical properties and the rate of evolution, necessary
to relate them to the local distribution of clusters of
galaxies. Large separation quasars are probably a good alternative way
to detect high redshift structures.

\acknowledgements
We wish to acknowledge fruitful discussion with D. Clowe, Y. Mellier and
P. Schneider.  JPK acknowledges support from CNRS, JH from the Danish
Natural Science Research Council (SNF).


\end{document}